# A Preliminary Investigation in the Molecular Basis of Host Shutoff Mechanism in SARS-CoV

Niharika Pandala
Computer Science and Engineering
U of SC
npandala@email.sc.edu

Casey A. Cole
Computer Science and Engineering
U of SC
coleca@email.sc.edu

Devaun McFarland
Computer Science and Engineering
U of SC
mcfarlad@email.sc.edu

Anita Nag
Division of Natural Sciences and Engineering
USC Upstate
anitan@uscupstate.edu

Homayoun Valafar
Computer Science and Engineering
U of SC
homayoun@cse.sc.edu

**ABSTRACT**

Recent events leading to the worldwide pandemic of COVID-19 have demonstrated the effective use of genomic sequencing technologies to establish the genetic sequence of this virus. In contrast, the COVID-19 pandemic has demonstrated the absence of computational approaches to understand the molecular basis of this infection rapidly. Here we present an integrated approach to the study of the nsp1 protein in SARS-CoV-1, which plays an essential role in maintaining the expression of viral proteins and further disabling the host protein expression, also known as the host shutoff mechanism. We present three independent methods of evaluating two potential binding sites speculated to participate in host shutoff by nsp1. We have combined results from computed models of nsp1, with deep mining of all existing protein structures (using PDBMine), and binding site recognition (using msTALI) to examine the two sites consisting of residues 55-59 and 73-80. Based on our preliminary results, we conclude that the residues 73-80 appear as the regions that facilitate the critical initial steps in the function of nsp1. Given the 90% sequence identity between nsp1 from SARS-CoV-1 and SARS-CoV-2, we conjecture the same critical initiation step in the function of COVID-19 nsp1.

## 1. Introduction

In light of the worldwide spread of the novel coronavirus disease (COVID-19), our ability to harness the technological advances acquired over the past 2-3 decades is being challenged. Encounter with COVID-19 has demonstrated our ability to rapidly and cost-effectively isolate the genome sequence of the new emerging pathogen, such as SARS-CoV-2. However, we have also encountered significant impediments in the ability to translate this knowledge into a tangible mechanism of combating new biological threats. New technologies need to develop to convert the genomic information into protein structure/function to eliminate this translational operation gap.

During the CoVID-19 outbreak, much of the attention has been focused on spike protein and its binding to the lung cell receptor ACE-2,[1] while very little is known about the way the novel coronavirus (HCoV-19 or SARS-CoV-2) reprograms the normal cellular function. SARS-CoV-2 shares a high degree of homology to severe acute respiratory syndrome coronavirus (SARS-CoV) that precipitated an outbreak in 2003, resulting in a 9.6% fatality.[2] The sequence homology of two viruses allows us to compare the structure and function of specific proteins by extending our understanding of SARS-CoV to SARS-CoV-2. SARS-CoV-2, a positive-strand RNA virus, produces sixteen nonstructural proteins that facilitate viral propagation. Out of these sixteen proteins, nonstructural protein 1 (nsp1) suppresses the host's immune response and also acts as the host shutoff factor and dampens host gene expression. Nsp1 triggers selective decay of host messenger RNAs (mRNA) by stalling ribosome assembly and cleaving the host mRNA. The study of nsp1 offers a significant challenge due to its unique sequence and disordered structure, making it difficult to study its structure-function relation by traditional methods.

The traditional characterization of protein structures by NMR spectroscopy or X-ray crystallography is too time-consuming and too expensive. It does not apply to challenging proteins such as intrinsically disordered proteins. The computational modeling tools that have emerged in the recent decade have continued to evolve with evaluation in venues such as the Critical Assessment of Techniques for Protein Structure Prediction (CASP) competition. In this exploration, we propose to combine several computational approaches and integrate them with experimental validation to develop rapid deciphering of the protein interactions of COVID-19. More specifically, we aim to decrypt the elusive function of the nsp1, a protein with an intrinsically disordered region, in the family of coronavirus by expanding our current knowledge from SARS-CoV to SARS-CoV-2.

## 2. Background

The virus causing Severe Acute Respiratory Syndrome (SARS) belongs to the coronavirus family. SARS coronavirus (SARS-CoV) is a zoonotic virus that has crossed the species barrier from bats to humans by an unknown mechanism. This virus emerged from Guangdong, China, in 2002[3] and quickly spread across the globe. The most recent outbreak of COVID-19 emanated from Wuhan, China, and continues to spread globally. The virus causing COVID-19, SARS-CoV-2, shares a sequence homology of 79.5% with SARS-CoV[4]; therefore, our investigations have a direct impact on better understanding its functional mechanism.

SARS-CoV contains a single-stranded RNA genome that replicates in the cytoplasm of host cells[3]. The replicated gene that includes two-thirds of the genome encodes two large precursor polyproteins that are cleaved into 16 nonstructural proteins (the nsp family of proteins), among which nsp1 is the first to be translated[4]. Nsp1 causes severe translational shutdown of host proteins from host mRNAs, leads to suppression of host gene expression. This 180-residue protein consists of two unstructured domains and one structurally well-defined domain. The two unstructured domains of nsp1 consist of residues 1-12 and residues 126-180, while the structurally well-defined region consists of the interconnecting region (residues 13-126). Residues 13-126 form a well-defined structure that has been previously studied by NMR spectroscopy (see Figure 1)[5]. Residues 1-12 and 127-180 constitute poorly characterized regions, difficult to be examined by traditional methods due to their dynamic properties.

Laboratory-based mutation of this protein has demonstrated both amplifying and attenuating activities on its host shutdown properties.[6] Many of these mutations are concentrated on the surface residues of nsp1; therefore, the structural implication of these mutations may provide a better understanding of the protein's ability to dampen host gene expression. In this study, we have explored two potential regions of interest shown in Table 1 that alter nsp1's ability to shut down host gene expression. To better understand the structural basis of nsp1's function, we integrate advanced protein structure modeling I-TASSER[7], multiple structure alignment technique msTALI[8-10], and deep mining of the Protein DataBank using PDBMine[11].

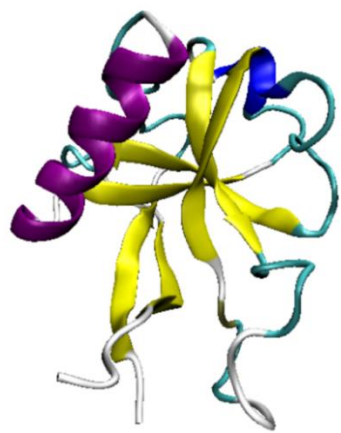

Figure 1. Structured region of Nsp1 determined NMR spectroscopy (PDB ID: 2GDT[12] and 2HSX[12]).

Table 1. Laboratory confirmed mutations that inhibit host gene expression by amplifying nsp1's function.

| Gene Expression Inhibited by 150-450% |
|---|
| E55R, E57R, K58E, G59R |
| R73E, D75R, L77A, S78E, N80G |

## 3. Materials and Methods

Our investigation of the structure-function relationship of nsp1 is based on the mutations outlined in Table 1. While the functional effects of these genomic alterations have been investigated, their corresponding structural effects are mostly unknown. Therefore, we employed a protein structure modeling technique to investigate these effects and close the knowledge gap economically and efficiently. Our overall approach consists of three independent investigations of the functionally important regions of nsp1. By combining the results of independent approaches, a consensus of results is drawn to identify the most likely region of its function. The details of our approach are described in the following sections.

### 3.1. Computational Protein Modelling

Protein modeling software I-TASSER[7] was utilized for the structural modeling of the mutations to nsp1 to understand the mechanism of each mutation better. I-TASSER was selected due to its continued high performance during the CASP competitions[13]. As a preliminary step, we first evaluated the reliability of I-TASSER[7] by comparing the modeled structure of wild-type nsp1 to the experimentally characterized structure of nsp1 in the Protein Data Bank[14] (PDB-ID: 2GDT[12] and 2HSX[12]). As a second step, we used I-TASSER to generate the structure of reported mutants of nsp1 in Table 1. These structures were then used to perform correlative studies between the observed structural changes and reported functional changes. In particular, we examined the steric and charge distribution of the examined mutations to establish the structural basis of changes in function.

### 3.2. Visualization and Calculation of Energies and Surface Perturbation

All the visualizations, calculations of energies, and quantification of surface perturbation were performed using VMD[15] and NAMD[16] (Nanoscale Molecular Dynamics). A solvation box was introduced in the calculation of NAMD energies for each protein. The solvent being water at a pH of 7.4. This approach of including water in energy simulations produces a more realistic physiological environment for the protein and therefore produces more viable results. The analysis protocol consists of subjecting the solvated structural models of each mutant to 1,000 rounds of molecular mechanics energy minimization, followed by 100,000 rounds (0.1ns total simulation time) of molecular dynamics simulation using NAMD [16].

The topological variation on nsp1's surface compared to its mutations was studied by measuring the sidechain length of the substituted amino acids. Then quantified the perturbation of the surface by measuring the changes in the volumetric occupancy of the mutated amino acid side chains. The distance used in this calculation were measured between the backbone $C_\alpha$ atom of mutation and the furthermost atom in space and then compared to the respective region in the wild-type nsp1 protein.

### 3.3. Mining of PDB with PDBMine

PDBMine[11] is a recently introduced curation of all protein structures from PDB[17] to facilitate extensive mining of structural information. PDBMine was created by dissecting and aggregating all existing structural information such as sequences, backbone dihedrals, water accessibility of each residue (to name a few) of every characterized protein structure. Mining of the PDBMine can quickly produce the conventional Ramachandran plots[18], and more extensive correlative knowledge, such as the relationship between dihedrals between all sets of adjacent residues. PDBMine has also facilitated the discovery of absent k-mer sequences[19, 20] in the PDB.

In this study, we have utilized PDBMine to discover any existing protein structures with similar k-mer sequence to SARS-CoV nsp1 protein. More specifically, 174 6-mers were created from the primary sequence of nsp1 using a rolling window size of 6 residues (e.g., residues 1-6 in window 1, 2-7 in window 2, etc.). The same exercise was repeated with varying window sizes 4-7, and the most informative window sizes of 6 and 7 were analyzed. In theory, a smaller window size will discover a larger number of proteins with a less significant relationship. Therefore, results obtained from the largest window sizes (6 and 7 in this instance) have the most significance in structural inference. It is important to note that a random (unrelated) similarity of a 6 or 7-mer will appear only once in 6.4e7 or 1.28e9, respectively. Thus, any discovered similarity resulted from the search of only 140,000 proteins is of significant interest. The use of PDBMine in this context allows the discovery of a significant relationship between nsp1 and other existing proteins to infer structural similarity. Such structural similarities can be used to construct the protein's fragmented structure or assist in the discovery of the binding sites[9].

### 3.4. Binding Site Recognition with msTALI

Identification of the active site in proteins can be an essential step in understanding the mechanism of its function. Typical approaches to identification of an active site are based on the docking of the protein with its substrate. In this approach, the "fitness" compatibility of the protein and its substrate is evaluated based on known physical interactions such as charge, shape, or hydrophobic/hydrophilic compatibility between the suspected sites of a protein and its known substrate. While such approaches are complete, they impose a computationally demanding task since they must explore all possible orientations and conformations of the substrate. msTALI[10] presents a new paradigm for identifying an active site based on the evolutionary relationship between a group of proteins with known similar substrate binding properties. Under this paradigm, msTALI and its predecessor TALI[21, 22] can identify the active site based on structural and sequence homology of multiple proteins. msTALI has demonstrated unparalleled success in the identification of active sites such as ATPase[9], P-loop folds of the G proteins, PYP-like in Profilin-like folds, and FAD-linked reductases family in FAD/NAD(P) folds.[8] In this investigation, a select group of proteins (shown in Table 2) were subjected to a study by msTALI. The selection of these proteins was based on the relevance of their function to the binding of nsp1, CATH class diversity, and size diversity to provide a diverse group of proteins. The alignment of this collection of proteins should provide an independent method of obtaining the conserved binding site.

Table 2. The list of proteins with related functional activity as nsp1 that were used with msTALI.

| PDB-ID | Size | CATH |
|---|---|---|
| 1HUS | 155 | 1.10.455.10 |
| 1IQV | 218 | 1.10.455.10 |
| 1VI6 | 208 | 3.40.50.10490 |
| 1QKH | 92 | 3.30.860.10 |
| 1QXF | 66 | 2.20.25.100 |
| 1RIP | 81 | 2.40.50.140 |
| 1RSS | 151 | 1.10.455.10 |
| 2FKX | 88 | 1.10.287.10 |
| 2MEW | 82 | 3.30.70.600 |
| 3BN0 | 112 | 3.30.1320.10 |
| 4BSZ | 244 | 3.30.300.20\| 1.25.40.20\| 3.30.1140.32 |
| 6G04 | 156 | NA |
| NSP1 | 180 | NA |

## 4. Results and Discussion

In the following sections, we provide the results of our investigation to elucidate the functionally important regions of nsp1 and to describe a hypothesized mechanism of its function better. We start our investigation through the most traditional explanation of structure-function relationship and end with utilizing contemporary Bioinformatics approaches. The conventional investigations consist of exploring the destabilizing effects of the mutations by observing potential energy, steric hindrance, and the change in charge distribution of each mutation. Our contemporary approaches leverage the current information regarding other similar proteins to assist with the discovery of the conserved and functionally important regions while forming a more complete mechanism of molecular interactions of nsp1.

### 4.1. Predicted Structures

As a preliminary step, we first evaluated the reliability of I-TASSER[7] by comparing the modeled structure of wild-type nsp1 to the experimentally characterized structure of nsp1 in the Protein Data Bank[14] (PDB-ID: 2GDT[12]). A comparison of these two structures (shown in Figure 2) yielded a backbone structural difference (RMSD) of 1.95 Å, which is within an acceptable range of less than 3Å. Notably, the differences arise in the flexible loop region.

As the next step, we investigated the possibility of significant structural changes compared to the wild-type nsp1 due to the two sets of mutations mentioned in Table 1. The computed structures of nsp1 with the two different mutations from Table 1 are shown in Figure 2. Compared to the computed structure of nsp1, both structures exhibited structural similarity under 2Å, denoting no significant perturbation of structure that could explain the disruption of its function. The primary conclusion is to focus our investigation on other explanations for the functional perturbation of nsp1.

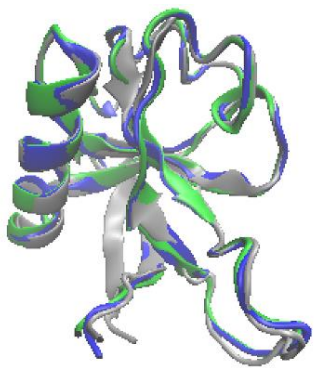

Figure 2. A superimposition of wild-type nsp1 (grey) with mutation E55R, E57R, K58E, G59R (blue) and mutation R73E, D75R, L77A, S78E, N80G (green).

### 4.2. Effect of Charge Distribution

The charge interactions were studied at a neutral pH value of 7.4 to understand the polarity and the nature of change in the substitution of amino acids. Both sets of mutations shown in Table 3 are present in the surface-exposed region of nsp1 but are spatially separated, presumably interacting with different sets of host factors (Figures 3 and 4). In these figures, mutated amino acids are color-coded: acidic amino acids in red, alkaline amino acids in blue, and neutral amino acids in yellow.

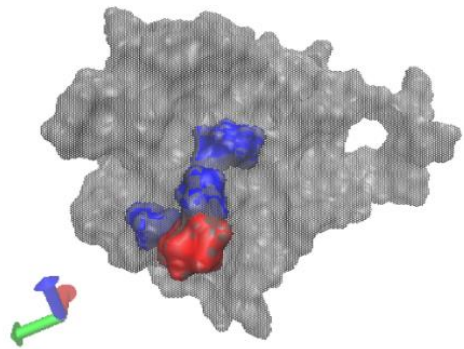

Figure 3. Charge interactions of E55R, E57R, K58E, G59R mapped on wild-type nsp1.

The mutation E55R, E57R, K58E, G59R is present as clustered on the surface with an outward protrusion on the surface of nsp1 (Figure 3). Upon mutation, the overall change in amino acids from acidic (E: Glutamic acid) to alkaline (R: Arginine) maintains the hydrophilicity of the surface residues but exposes positively charged alkaline amino acids on the surface. Since nsp1 binds to mRNA and the ribosome, a positively charged side chain may facilitate interaction with the negatively charged backbone of RNA.

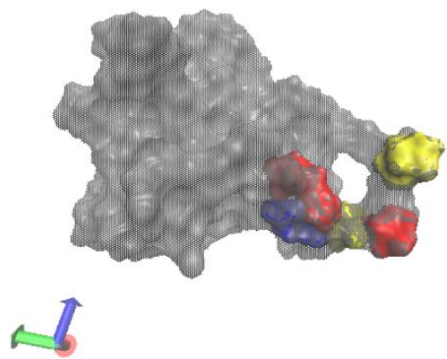

Figure 4. Charge interactions of R73E, D75R, L77A, S78E, N80G mapped on wild-type nsp1.

The mutation R73E, D75R, L77A, S078E, N80G is present around an extended loop on nsp1 protein (Figure 4). In contrary to the 55-59 region, mutations in 73-80 region changes amino acid residues to primarily acidic residues.

NAMD energy calculations in the solvated state (Table 3) show that the two mutations have resulted in significantly different potential energies. Together, these data suggest each set of surface residues function independently to bind the host proteins and facilitate host shutoff.

Table 3. NAMD energy calculations are shown for the wildtype protein as well as the two mutants.

| Protein | NAMD energy Kcal/mol |
|---|---|
| wild-type nsp1 | -12,609.9 |
| E55R, E57R, K58E, G59R | -16,038.1 |
| R73E, D75R, L77A, S78E, N80G | -10,313.5 |

### 4.3. Effect of Surface Perturbation

The surface perturbation of the two mutant structures was measured using the VMD software suite (visualizations shown in Figure 5 and Figure 6). The change in the sidechain occupancy of each amino acid substitution then compared to the distances seen in the wild-type protein. These results are summarized in Table 4 and Table 5. Each column in these tables represents a given amino acid substitution in the mutant structures and shows the observed topological changes when compared to the computed structure of the wild-type nsp1. All the structural changes are due to the extended sidechains of the substituted amino acids that protrude into the water accessible region of the protein. These perturbations of surface properties can provide a steric-based mechanism of enhancing nsp1's binding or recognition function; therefore, amplifying its host shutoff function.

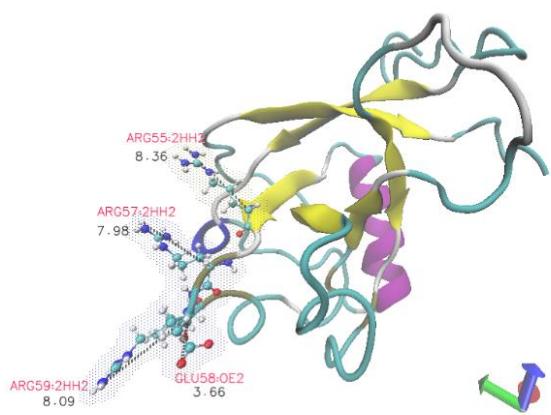

Figure 5. VMD bond length calculations for nsp1 mutant E55R, E57R, K58E, G59R.

Table 4. Results of the quantification of surface perturbation for mutations E55R, E57R, K58E, G59R

| Amino Acid Substitution | Distance change Compared to wild-type (Å) |
|---|---|
| E55R | Increases (5.62) |
| E57R | Increases (3.61) |
| K58E | Decreases (-1.99) |
| G59R | Increases (5.70) |

For the nsp1 mutant E55R, E57R, K58E, G59R, all the substitutions form charged hydrophilic nodes. Arginine (R) substitution contributes to long-range connectivity, hence increases bond lengths. These long-range interactions may facilitate the cooperative stabilization of the native conformation[23, 24], thus, altering the binding properties of the binding site.

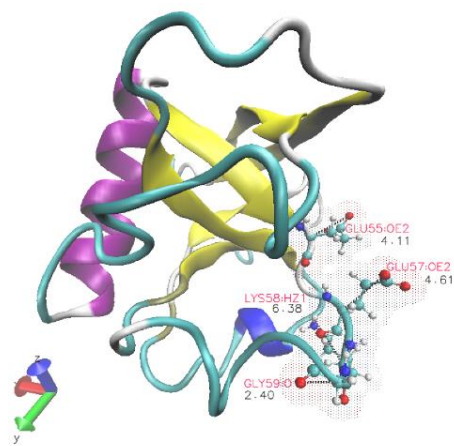

Figure 6. VMD bond length calculations for nsp1 mutant R73E, D75R, L77A, S78E, N80G.

Table 5. Results of the quantification of surface perturbation for mutations R73E, D75R, L77A, S78E, N80G

| Amino Acid Substitution. | Distance change Compared to wild-type (Å) |
|---|---|
| R73E | Decreases (-2.97) |
| D75R | Increases (4.43) |
| L77A | Decreases (-1.88) |
| S78E | Increases (1.48) |
| N80G | Decreases (-1.77) |

The nsp1 mutant R73E, D75R, L77A, S78E, N80G is a combination of hydrophilic (charged amino acids: E, R) and neutral (small nonpolar amino acids: A, G) side chains. Three out of five mutations are hydrophilic, and most mutations have a relatively small difference in sidechain length compared to the wild-type protein. They form short-range interactions around the external loop has the potential for a highly cooperative transition to interact with mRNA or ribosome to facilitate suppression of host gene expression.

### 4.4. Predicted Binding Site by msTALI

The results of the structural alignment of the thirteen proteins listed in Table 2 using msTALI are shown in Figure 7. In this figure, a higher score implies a higher significance of the conserved region. The larger spikes shown in this figure clearly indicate four significant regions that are shown in Table 6. There are two particularly interesting observations. The first clearly indicates the importance of the region 74-80, which is one of the implicated functional regions of nsp1. The second observation relates to the near-complete absence of any reported significance for the region 55-60. In summary, in the context of structural similarity to other proteins with relevant binding properties, the region 74-80 is the more likely initial binding site to the rRNA complex. Based on this result, it can be hypothesized that the three critical and buried conserved regions are the requisite structural requirement to form the structure of this protein and facilitate the docking process at residues 74-83.

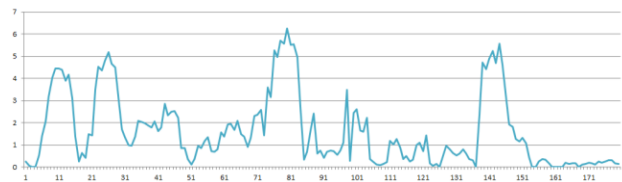

Figure 7. The msTALI conservation score for SARS-CoV-1 nsp1 protein with respect to the other proteins. A higher score is an indication of a more significant conserved region.

Table 6. The conserved regions across the 12 proteins with related functions to that of nsp1.

| Name | Range | Surface Accessibility |
|---|---|---|
| Region 1 | 7-15 | Partial surface exposure |
| Region 2 | 22-29 | Partial surface exposure |
| Region 3 | 74-83 | Complete surface exposure |
| Region 4 | 139-146 | Minimal surface exposure |

### 4.5. Results of the PDBMine Search

PDBMine was used with rolling window sizes of 3, 4, 5, 6, and 7. Generally, a larger window size will result in the identification of a more significant relationship. Therefore, here we present the results for a window size of 7, although a window size of 6 also presented some interesting relationships. Table 7 summarizes the most notable regions for a PDBMine search of a window size of 7. The first and second columns of this table represent the region and the corresponding sequence of nsp1, respectively, with which other proteins shared identity. The last column indicates the number of proteins that contained the exact sequence listed in the second column. Although not explored here, further analysis of these particular proteins can provide critical information to provide additional insight into the function of nsp1. For instance, in our investigation, there is substantial similarity between the findings of msTALI (in Table 6) and PDBMine search including the highlighted region of 70-76. Of additional interest, nsp1 exhibits a 7-mer identity starting at residue 167 (SGALREL) with proteins 5JHQ and 4YSB. This segment of a sequence exhibits the conserved helical structure shown in Figure 8. Based on our initial results, this helical region may be the only structured region in the entire unstructured region of nsp1 (residues 128-180) and may play a significant role in further recognition of its substrate.

Table 7. Results of PDBMine search with a window size of 7.

| Residue Range | Sequence | Number of hits |
|---|---|---|
| 23-29 | VRDVLVR | 14 |
| 35-42 | VEEALSEA | 25 |
| 70-76 | FIKRSDA | 52 |
| 137-143 | GIDLKSY | 2 |
| 167-173 | SGALREL | 2 |

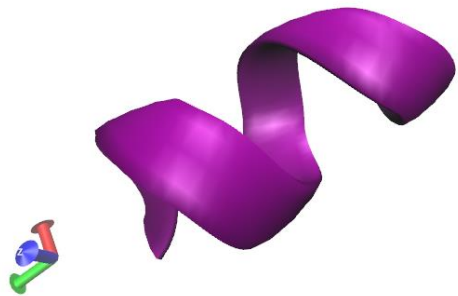

Figure 8. The conserved structural region shared by 5JHQ and 4YSB.

## 5. Conclusion and Future Direction

The traditional study of structure-function relation of nsp1 by investigating energetics suggests that these two mutations function by distinctly different mechanisms. The mutations in the 55-59 region stabilize the protein, while the mutations in the 73-80 region destabilize the protein. Since flexibility is the key feature of nsp1, each mutation can function differently in the presence of host factors (RNA and proteins).

While both mutations amplify the host shutoff function of nsp1, mutations to positively charged alkaline amino acids in the 55-59 region display an increase of side chain length that may favor long-range binding as well as interaction with nucleic acids. The mutations in the 73-80 region do not change the charge distribution or side chain length significantly but might support binding host factors due to its inherent flexibility.

The msTALI studies clearly identified the region 75-80 as the critical binding site in the context of the ribosomal binding function, while excluding the region 55-60. Furthermore, the region 72-80 was the only surface-exposed conserved region, which further supports this region's binding propensity. Finally, three other regions became identified as significantly conserved across all thirteen proteins. To speculated that these three buried regions indirectly support the function of nsp1 by forming the hydrophobic core of the nsp1, partial surface accessibility of some of these regions may also assist the critical interactions with the substrate of this protein.

The PDBMine search corroborated along with the results of msTALI, moreover it provided additional information that could pave the future of our investigations. There are two important contributions that require further investigation. The first relates to the 52 identified proteins that share the same 7-mer (residues 70-76) as nsp1. Examining the regions corresponding to this heptamer may provide additional information regarding the functional importance of this region starting at residue 70-76. The second area of investigation will focus on the short but stable helix (residues 167-173) identified in the unstructured region of nsp1. This short helical region may be the critical step in shaping the flexible region of nsp1 around the bound substrate.

## ACKNOWLEDGMENTS
This work was supported by NIH grant number P20 RR-016461 to Dr. Homayoun Valafar, BIPP funding from P20 RR-016461 to Dr. Nag, USC-RISE funding to Dr. Nag, and USCRF-VPR fund 155900-20-54157 to Drs Valafar and Nag.